\documentclass[manuscript]{aastex}

\usepackage{emulateapj5}

\newcommand{\Mj}{\,$M_{\rm Jup}$}

\slugcomment{Accepted for publication in the ApJ}

\shorttitle{A Methane Isolated Planetary Mass Object in Orion}
\shortauthors{Zapatero Osorio et al.}

\begin{document}

\title{A Methane Isolated Planetary Mass Object in Orion}
 
\author{M.\,R$.$ Zapatero Osorio\altaffilmark{1}, 
        V.\,J.\,S$.$ B\'ejar\altaffilmark{2}, 
        E.\,L$.$ Mart\'\i n\altaffilmark{3}, 
        R$.$ Rebolo\altaffilmark{2,4}, 
        D$.$ Barrado y Navascu\'es\altaffilmark{1},
        R$.$ Mundt\altaffilmark{5},
        J$.$ Eisl\"offel\altaffilmark{6},
    and J.\,A$.$ Caballero\altaffilmark{2}}

\altaffiltext{1}{LAEFF-INTA, P.\,O$.$ Box 50727, E-28080 Madrid, Spain; 
                 mosorio@laeff.esa.es, barrado@laeff.esa.es}
\altaffiltext{2}{Instituto de Astrof\'\i sica de Canarias, E-38205 
                 La Laguna, Tenerife, Spain; vbejar@ll.iac.es, 
                 rrl@ll.iac.es, zvezda@ll.iac.es}
\altaffiltext{3}{Institute of Astronomy, Univ. of Hawaii at Manoa, 2680 
                 Woodlawn Drive, Honolulu, HI 96822 USA; 
                 ege@ifa.hawaii.edu}
\altaffiltext{4}{Consejo Superior de Investigaciones Cient\'\i ficas,
                 Madrid, Spain}
\altaffiltext{5}{Max-Planck-Institut f\"ur Astronomie, K\"onigstuhl 17, 
                 D-69117 Heidelberg, Germany; mundt@mpia-hd.mpg.de}
\altaffiltext{6}{Th\"uringer Landessternwarte Tautenburg, Sternwarte 5, 
                 D-07778 Tautenburg, Germany; jochen@tls-tautenburg.de}

\begin{abstract}
We report on the discovery of a free-floating methane dwarf toward the
direction of the young star cluster $\sigma$\,Orionis. Based on the
object's far-red optical and near-infrared photometry and
spectroscopy, we conclude that it is a possible member of this
association. We have named it as S\,Ori\,J053810.1--023626 (S\,Ori\,70
is the abridged name). If it is a true member of $\sigma$\,Orionis,
the comparison of the photometric and spectroscopic properties of
S\,Ori\,70 with state-of-the-art evolutionary models yields a mass of
3\,$^{+5}_{-1}$ Jupiter mass for ages between 1\,Myr and 8\,Myr. The
presence of such a low-mass object in our small search area
(55.4\,arcmin$^2$) would indicate a rising substellar initial mass
function in the $\sigma$ Orionis cluster even for planetary masses.
\end{abstract}


\section{Introduction}

Since the discovery of brown dwarfs (i.e., objects unable to burn
hydrogen stably in their interiors and with masses below 72\Mj; Kumar
\cite{kumar63}; Chabrier et al$.$ \cite{chabrier00}) both in the field
and in young open clusters (see Basri \cite{basri00} for a review),
many questions remain unsolved. A very important one is the minimum
mass for the formation of very low mass objects in isolation, which
would represent the bottom end of the initial mass function (IMF) for
free-floating objects. B\'ejar et al$.$ \cite{bejar01} have
scrutinized the substellar population of the young, nearby $\sigma$
Orionis cluster, and derived a rising IMF
(d$N$/d$M$\,$\sim$\,$M^{-0.8\pm0.4}$), which is complete for brown
dwarf masses in the range of 72--13\Mj. Other authors find similar
substellar IMFs (e.g., Luhman et al$.$ \cite{luhman00}). Very recent
photometric and spectroscopic searches (Lucas \& Roche \cite{lucas00};
Zapatero Osorio et al$.$ \cite{osorio00}) suggest that the IMF extends
further below the deuterium burning mass threshold at around
13\Mj~(Saumon et al$.$ \cite{saumon96}; Burrows et al$.$
\cite{burrows97}). We will refer to this mass regime as the
``planetary mass'' domain. The least-massive objects so far identified
in young stellar clusters of Orion have masses around 5--10\Mj~(Lucas
et al$.$ \cite{lucas01}; Mart\'\i n et al$.$ \cite{martin01}), and
cover the full range of the spectral type L. For the formation
scenarios of such objects it would be very important to see down to
which masses they can be found in isolation.

Here we report on the discovery of a methane dwarf toward the
direction of Orion. In this paper we present evidence for its
membership in the $\sigma$\,Orionis star cluster, which implies that
this object is likely the least massive planetary mass body imaged to
date outside the solar system. Cluster membership is discussed in
Section \ref{membership}, and mass determination based on
state-of-the-art evolutionary models and further discussion are
presented in Section \ref{mass}.

\section{Observations \label{obs}}

Our candidate was selected from a $J$, $H$ near-infrared survey, in
which the south-western region of the young $\sigma$\,Orionis cluster
was targeted down to 3-$\sigma$ detection limit of $J, H \sim$
21\,mag.  An area of 55.4\,arcmin$^2$ was covered with the
near-infrared camera INGRID mounted at the Cassegrain focus of the
4.2-m William Herschel Telescope (WHT; at Roque de los Muchachos
Observatory).  This camera is equipped with a 1024$\times$1024 Hawaii
detector, which provides a pixel size of 0\arcsec.242 projected onto
the sky.  The observations were carried out on 2000 November 5--6;
these nights were photometric, but the seeing was quite poor at around
1\arcsec.8.  The total integration time was 3240\,s in each of the $J$
and $H$ filters. A nine-position dither pattern, repeated three times,
was used to obtain the images; each image consists of 4 co-added
exposures of 30\,s.  Standard data reduction, which included dark
subtraction and flat-fielding, was done under the {\sc
iraf\footnote{IRAF is distributed by National Optical Astronomy
Observatory, which is operated by the Association of Universities for
Research in Astronomy, Inc., under contract with the National Science
Foundation.}}  environment. Instrumental magnitudes were converted
into observed magnitudes using short integration observations of the
INGRID fields obtained with the near-infrared camera available at the
1.5-m Carlos S\'anchez Telescope (Teide Observatory) on the
photometric nights of 2002 April 2--3. UKIRT standard stars taken from
the work of Hunt et al$.$ \cite{hunt98} were observed throughout this
night.

\begin{figure}
\plotone{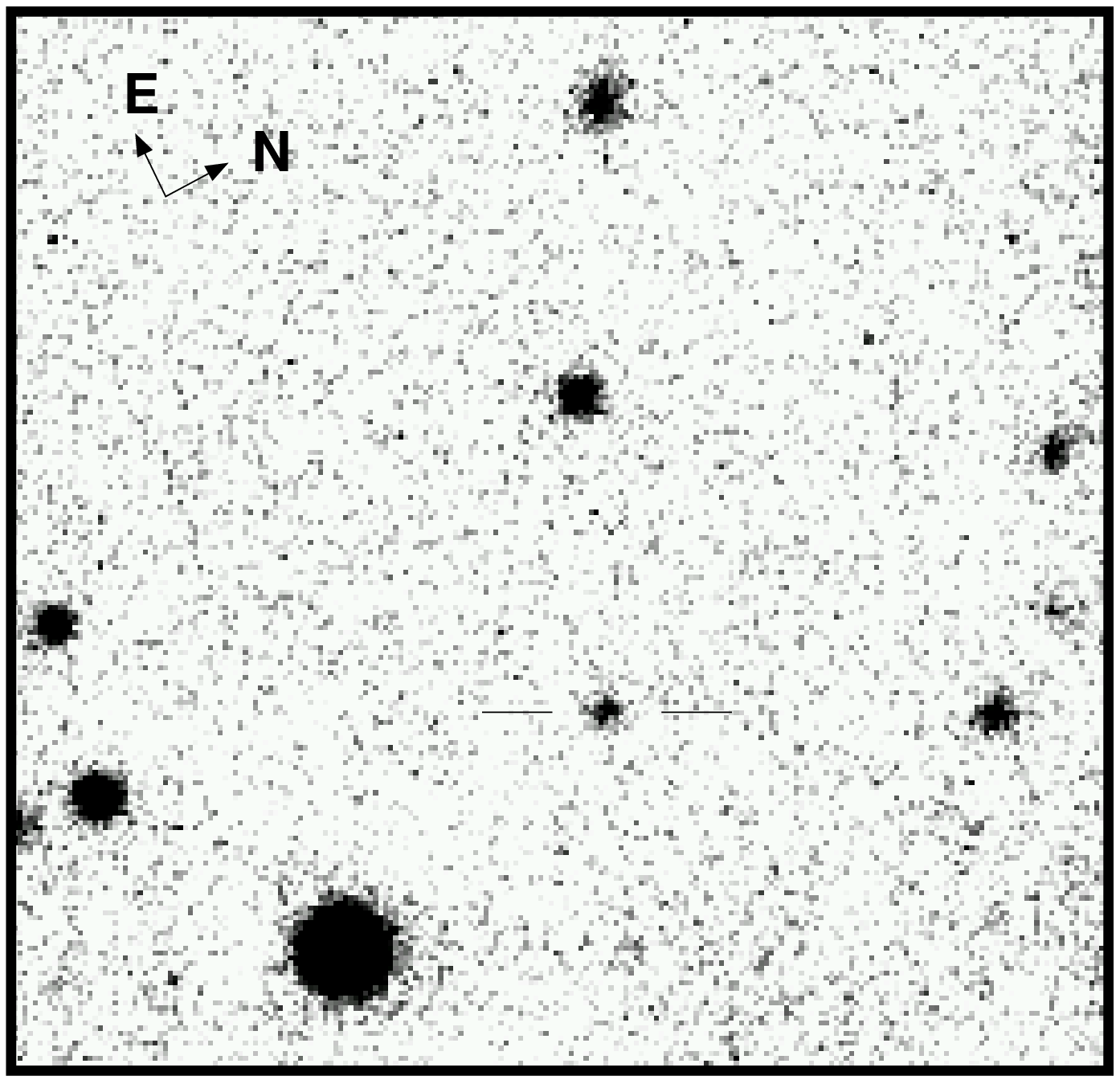}
\caption{S\,Ori\,70 finder chart (NIRC $K_s$ image, 
         32\arcsec$\times$32\arcsec). \label{fc}}
\end{figure}

Among various photometric candidates for possible membership in
$\sigma$\,Orionis, one showed a rather blue $J-H$ color of about
--0.1\,mag and $J$\,=\,20.28\,mag. We inspected previous $I$- and
$z$-band images obtained with the Low Resolution Imaging Spectrometer
(LRIS; Oke et al$.$ \cite{oke95}) at the 10-m Keck II telescope on
1998 December 23 to identify this INGRID candidate -- the INGRID
survey mostly overlapped with the LRIS optical images. It appeared at
the detection limit with $I$\,=\,25.0\,mag and
$I-z$\,=\,2.2\,mag. LRIS is equipped with a Tek 2048\,$\times$\,2048
CCD detector, which provides an imaging scale of
0\arcsec.215\,pix$^{-1}$. Our optical LRIS survey covered an area of
218.4\,arcmin$^2$ in the $\sigma$\,Orionis cluster. Individual
exposures of 5\,min integration time were taken in each filter; the
average seeing was 0\arcsec.7--0\arcsec.8. $I$-band magnitudes were
calibrated with respect to previously known $\sigma$\,Orionis brown
dwarfs, like S\,Ori\,47 (Zapatero Osorio et al$.$
\cite{osorio99a}). The $z$-band data were calibrated as in Zapatero
Osorio et al$.$ \cite{osorio99b}; this is not an absolute calibration
and is relative to the color $I-z$, which is supposed to be 0.0 for
A0-type stars. Further details on this LRIS search will be given
elsewhere (B\'ejar et al$.$ \cite{bejar02}). Based on the $IzJH$
colors, our candidate turns out to be an extremely red object. We have
named it as S\,Ori\,70; IAU nomenclature and coordinates (accurate up
to 2\arcsec) are given in Table~\ref{table}. Figure~\ref{fc} displays
a finder chart for S\,Ori\,70.

$K_s$-band photometry and $HK$ low-resolution grism spectroscopy have
been collected for S\,Ori\,70 with the NIRC camera (256$\times$256
pixel InSb detector; Matthews \& Soifer \cite{matthews94}) at the 10-m
Keck I telescope on 2001 December 29--30. Meteorological conditions
were photometric, and the seeing was around 1\arcsec~at the time of
the spectroscopic observations. The $K_s$-band images consisted of
five co-added exposures of 15\,s each. The spectra were obtained with
three co-added exposures of 150\,s each using the $HK$ filter, the
grism gr120 and a slitwidth of 3.5\,pixel. This configuration provides
simultaneous coverage of the $H$- and $K$-band region from
1.44\,$\mu$m to 2.5\,$\mu$m with a dispersion of
5.95\,nm\,pixel$^{-1}$ and a resolution of $R$\,$\sim$\,90. Nine- and
three-position dither patterns were used for imaging and spectroscopy,
respectively.  Telluric absorption was removed using the NIRC spectrum
of the G0 star SAO\,112561, observed at the same airmass as
S\,Ori\,70.  Photometric calibration in the $K_s$-band in the UKIRT
photometric system was done with observations of L- and T-type field
brown dwarfs taken from Leggett et al$.$ \cite{leggett02}.  We have
checked that our calibration of the $JHK$-bands and that of the 2MASS
survey are consistent within $\pm$0.05\,mag ($J$) and $\pm$0.09\,mag
($H$, $K_s$). This ensures a reliable comparison to the 2MASS data of
ultra-cool objects (Section~\ref{membership}).  Table~\ref{table}
provides all available photometry, and Figures~\ref{fotom}
and~\ref{spectrum} present color-magnitude diagrams and the final
smoothed $HK$-band spectrum of S\,Ori\,70.

In this spectrum S\,Ori\,70 presents broad strong absorption features
that we safely identify as water vapor and methane bands. Hence, our
near-infrared spectroscopy confirms the ultra-cool nature of our
candidate: it belongs to the recently defined ``methane'' spectral
class T (Burgasser et al$.$ \cite{burgasser02}; Geballe et al$.$
\cite{geballe02}).  The observed optical and near-IR colors are also
consistent with this classification. This is the first unambiguous
detection of a ``methane'' object toward the line of sight of the
$\sigma$\,Orionis cluster.  Mart\'\i n et al$.$ \cite{martin01}
claimed the detection of the methane bandhead at 2.2\,$\mu$m in
S\,Ori\,69 (tentatively classified as T0), which is 2.8\,mag brighter
in the $J$-band than S\,Ori\,70. S\,Ori\,70 displays, however,
stronger methane bands than S\,Ori\,69, supporting its cooler
nature. Using water and methane molecular indices defined for the
spectra of field T-dwarfs (Burgasser et al$.$ \cite{burgasser02};
Geballe et al$.$ \cite{geballe02}), we assign a spectral type of T5.5,
with an uncertainty of about one subclass, to S\,Ori\,70.  However, we
note that this classification may be unreliable (see next Section).

\section{Cluster membership \label{membership}}

Before we can attempt to determine the mass of S\,Ori\,70, we first
have to establish its membership in the young $\sigma$\,Orionis star
cluster, which is 1--8\,Myr old (Zapatero Osorio et al$.$
\cite{osorio02}), and at a distance of 352\,pc (Perryman et al$.$
\cite{perryman97}). B\'ejar et al$.$ \cite{bejar01} provide further
discussion on other cluster properties. Our candidate was selected
because it follows the photometric sequence expected for the
$\sigma$\,Orionis cluster, as illustrated in
Figure~\ref{fotom}. Additionally, its near-infrared spectrum agrees
with our expectations for true cluster members: it is fainter and
cooler than the least luminous $\sigma$\,Orionis members so far
identified.

If S\,Ori\,70 did show high proper motion, this would point to it
being a very closeby object, and hence not a member of
$\sigma$\,Orionis.  Therefore, we have done astrometry on our images,
which have a maximum epoch difference of $\sim$3\,years (see
Section~\ref{obs}). The position of S\,Ori\,70 was measured relative
to the same set of four star-like sources (Figure~\ref{fc}) around the
target from one epoch to the next. Our measurement accuracy is limited
by the fact that S\,Ori\,70 is seen close to the detection limit in
our first epoch $I$- and $z$-band images. We find that its proper
motion is too small to be detected with the available data, and can
thus place an upper limit of 0\arcsec.1\,yr$^{-1}$ to its
motion. Comparing this value with the proper motions of low-mass stars
in the solar neighborhood (e.g., Phan-Bao et al$.$ \cite{bao01}), we
conclude that S\,Ori\,70 likely lies at a distance larger than
$\sim$40\,pc. If it were a field T6-type dwarf, it would be at
75\,$\pm$\,20\,pc; we have adopted the $JHK$ absolute magnitudes of
Gl\,229B (Leggett et al$.$ \cite{leggett02}; Nakajima et al$.$
\cite{nakajima95}), a well-known T6-type brown dwarf in the field.

Nevertheless, T-dwarfs with $J-K_s$\,$\le$\,0.6\,mag and
$I-J$\,$\ge$\,3.5\,mag are rather common in the field. The all-sky
surveys 2MASS and SDSS are finding nearby T-class brown dwarfs at a
space density of 0.0042--0.05\,pc$^{-3}$ (Burgasser et al$.$
\cite{burgasser02}; Tsvetanov et al$.$ \cite{tsvetanov00}; Leggett et
al$.$ \cite{leggett00}). The recent survey of Liu et al$.$
\cite{liu02} is consistent with these results. However, such values
are likely a lower limit to the true T-dwarf space density. Due to the
color selection criteria and the filters used by the searches, 2MASS
is especially sensitive to mid-T and late-T objects, whereas SDSS can
detect brown dwarfs in a wider temperature range, but at shorter
distances. D'Antona, Oliva \& Zeppieri \cite{dantona99} found three
``methane'' photometric candidates when inspecting ESO public deep
$IJK$ images, which covered 37\,arcmin$^2$ and were complete down to
$J$\,=\,23.1\,mag. One of them was spectroscopically confirmed by Cuby
et al$.$ \cite{cuby99}. We remark that the result of D'Antona et al$.$
\cite{dantona99} is expected from a substellar mass function that goes
as d$N$/d$M$\,$\sim$\,$M^{\alpha}$, where $\alpha$\,=\,--1. Correcting
for the appropriate area and limiting magnitudes, we obtain that
0.08--0.2 field T-dwarfs may be contaminating our INGRID survey.

Very recently, Chabrier \cite{chabrier02} has calculated the brown
dwarf IMF in the Galactic disk using state-of-the-art models for the
description of T-dwarfs, and normalizing the function at the bottom of
the main sequence. This author also included brown dwarf mass, age,
spatial, and binary probability distributions in his calculations, and
found that a mass function with a slope $\alpha$\,=\,--1 is probably
about the upper limit for a power-law fit of the IMF in the substellar
domain. Based on Chabrier's \cite{chabrier02} predictions for the
number of ``methane'' dwarfs from a substellar IMF described by the
afore mentioned power-law form, we estimate that $\sim$0.3 T-class
field brown dwarfs may appear in our survey. This contamination
(0.08--0.3 objects) is non-negligible, and might indicate that our
object is a field T dwarf. Nevertheless, we note that if the field
substellar mass function were better described by a power-law fit with
a lower index (e.g., Moraux, Bouvier, \& Stauffer \cite{moraux01};
Hillenbrand \& Carpenter \cite{hillenbrand00}) , or by a
lognormal-type function (see discussion in Chabrier
\cite{chabrier02}), the expected contamination of field ultra-cool
dwarfs would be smaller, making the cluster membership of S\,Ori\,70
more likely.

In addition to the analysis of the contamination by field T dwarfs in
our survey, we have compared our observed spectra of S\,Ori\,70 and of
the T6V field standard brown dwarf 2MASSI\,J0243137--245329 (Burgasser
et al$.$ \cite{burgasser02}) to a set of synthetic spectral energy
distributions, which cover a wide range of temperatures
(500\,$\le$\,$T_{\rm eff}$\,$\le$\,1800\,K) and gravities
(3.0$\le$\,\,log\,$g$\,$\le$\,5.0, in CGS units).  These theoretical
spectra, kindly provided by F$.$ Allard, are described in Allard et
al$.$ \cite{allard01}. Basically, we have adopted the {\sc ``cond''}
models, which treat dust as if condensed in deep layers of the
photosphere. It is believed that dust is no longer working as an
opacity source at near-infrared wavelengths in T-class atmospheres
(Tsuji, Ohnaka, \& Aoki \cite{tsuji99}; Allard et al$.$
\cite{allard01}; Burrows, Marley, \& Sharp \cite{burrows00}).
Observed and synthetic spectra were compared through a least square
minimization technique to determine the best-fitting $T_{\rm eff}$ and
log\,$g$ parameters for our two objects.  This is done by minimizing
the sum ($\sigma^{2}$) of the squares of the differences between the
flux values of the program objects and the theoretical models. We
looked for least squares matches based on the data sets in the
intervals of 1.50--1.69\,$\micron$ and 1.99--2.40\,$\micron$, which
include molecular bands from methane, water vapor, and carbon
monoxide.  Other wavelengths, with poorer signal-to-noise in their
fluxes and possibly affected by telluric absorption, were not
considered. Figure~\ref{logg} depicts $\sigma^{2}$ as a function of
the range of $T_{\rm eff}$ and log\,$g$ in the theoretical spectra.
We find that the best fit to the T6V 2MASS field standard brown dwarf
yields $T_{\rm eff}$\,=\,950\,$\pm$\,50\,K and
log\,$g$\,=\,5.0\,$\pm$\,0.5 (cm\,s$^{-2}$), in perfect agreement with
Gl\,229B (Marley et al$.$ \cite{marley96}; Allard et al$.$
\cite{allard96}; Oppenheimer et al$.$ \cite{oppenheimer98}).

The $HK$ spectrum of S\,Ori\,70, on the other hand, is better
reproduced by cooler and lower gravity models according to the least
square minimization technique. We find the best fits for $T_{\rm
eff}$\,=\,800\,$^{+200}_{-100}$\,K and log\,$g$\,=\,4.0\,$\pm$\,1.0
(cm\,s$^{-2}$). The observed spectra of S\,Ori\,70 and
2MASSI\,J0243137--245329, along with a few theoretical spectra are
shown in Figure~\ref{models}. The lower gravity of S\,Ori\,70 compared
to the field T6V dwarf again supports its youth.  Furthermore,
evolutionary models predict that, at the age of the $\sigma$\,Orionis
cluster, objects with similar temperatures and colors have logarithmic
gravities around 3.0--4.0, consistent with our determination.

The colors of S\,Ori\,70 fall within the range of measurements for T
dwarfs, as shown in Figure~16 of Burgasser et al$.$
\cite{burgasser02}. However, this object presents optical and
near-infrared colors that are somewhat different from those of field
dwarfs with late-T spectral types. We exclude the $z$-band from these
comparisons due to the lack of an absolute calibration for our
data. Table~\ref{color} lists differences with respect to the spectral
classes T6V, T7V, and T8V, which have $T_{\rm eff}$ similar to that of
S\,Ori\,70. Averaged 2MASS $JHK$ colors for the field have been taken
from Burgasser et al$.$ \cite{burgasser02}, and the $I$ color belongs
to Gl\,229B (T6V; Matthews et al$.$ \cite{matthews96}). Color
transformations between UKIRT and 2MASS photometric systems (Carpenter
\cite{carpenter01}) are significantly smaller than the relatively
large uncertainties in the photometry of S\,Ori\,70. We can explain
these differences in terms of low gravity. According to theoretical
computations, water absorptions in the near-infrared are stronger for
lower surface gravities, while methane absorptions appear weaker
(Allard et al$.$ \cite{allard96}).  At optical wavelengths, the
effects of gravity are noticeable in the strength of atomic lines
(essentially K\,{\sc i} and Na\,{\sc i}).  The second column of
Table~\ref{color} provides color differences for a model of 900\,K;
the high gravity model (log\,$g$\,=\,5.0) has been subtracted from the
low gravity one (log\,$g$\,=\,3.0).  Such dissimilarities are
comparable to those of S\,Ori\,70, supporting the object's low gravity
atmosphere.

On the basis of all the available photometric and spectroscopic
properties of S\,Ori\,70, in addition to the relatively low
contamination expected for T-class field interlopers ($\le$\,0.3
objects), we conclude that S\,Ori\,70 is a possible member of the
$\sigma$\,Orionis cluster.

\section{Mass determination and final remarks \label{mass}}

\begin{figure}
\plottwo{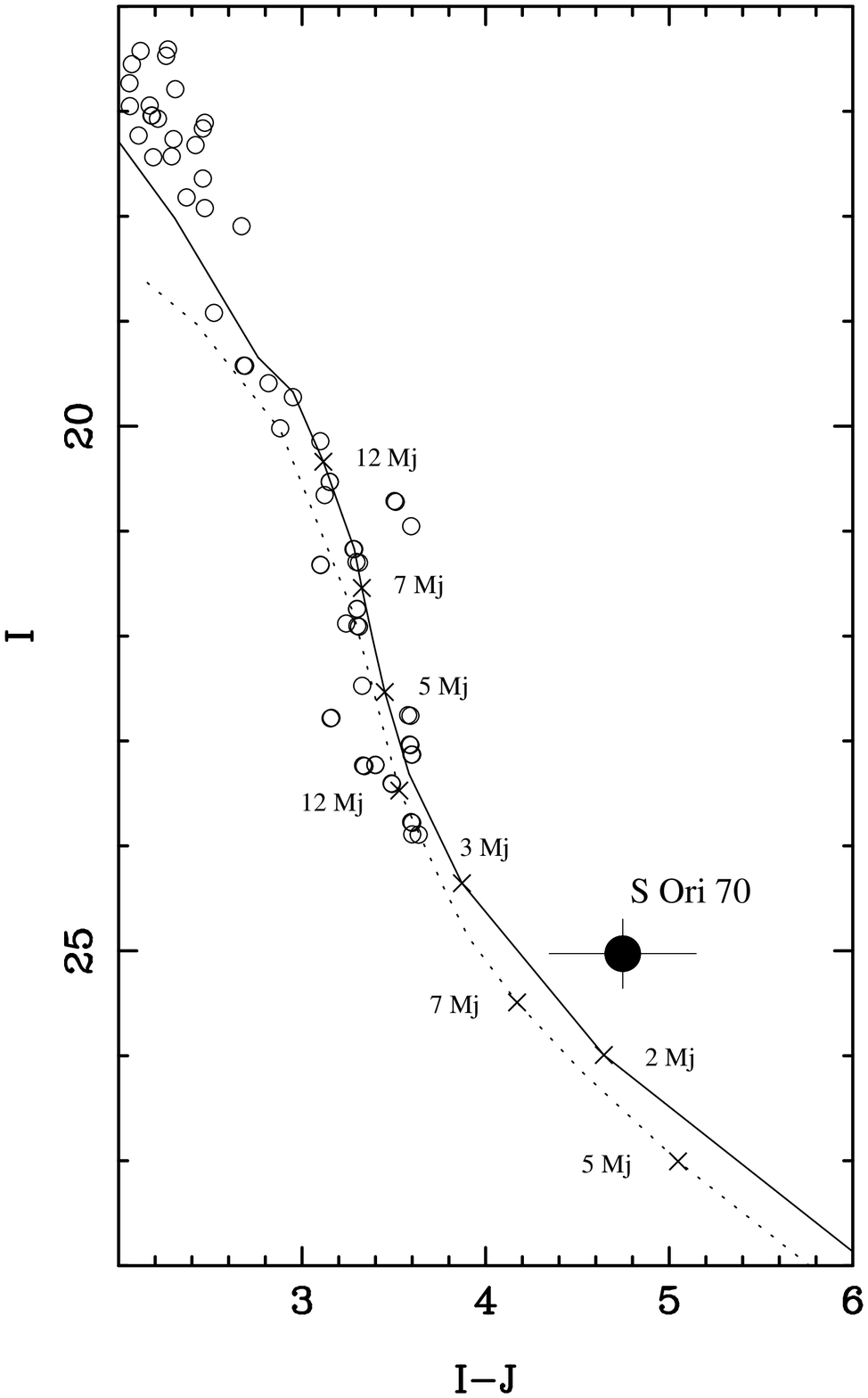}{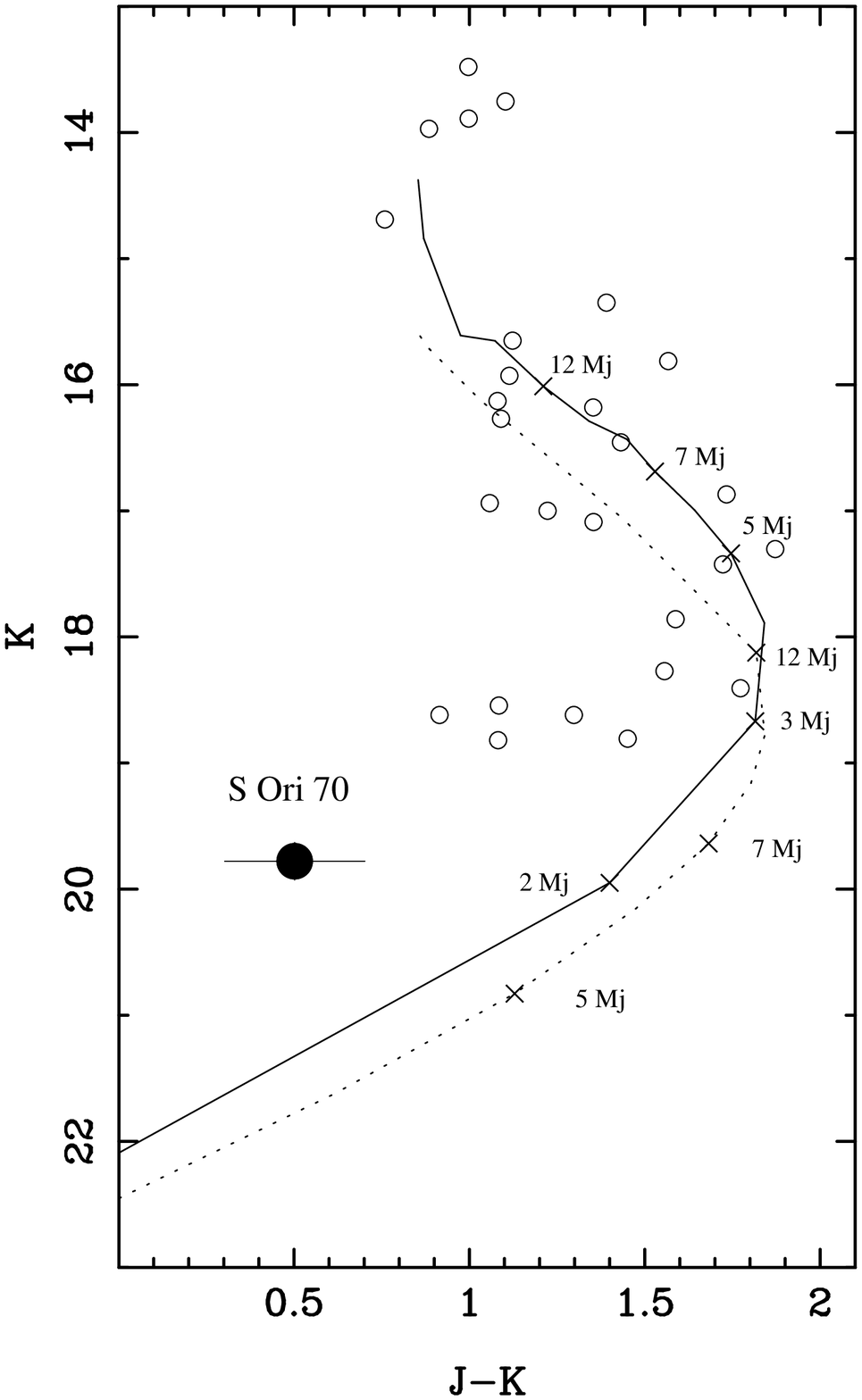}
\caption{S\,Ori\,70 extrapolates the optical and near-infrared
photometric sequence of the $\sigma$\,Orionis cluster; other cluster
members (B\'ejar et al$.$ \cite{bejar99}, \cite{bejar01}; Zapatero
Osorio et al$.$ \cite{osorio02}) are indicated with open
circles. Theoretical evolutionary models for 1\,Myr (solid line) and
10\,Myr (dotted line) are overplotted onto the data (Baraffe et al$.$
\cite{baraffe02}), and masses in units of Jupiter mass are also
provided. \label{fotom}}
\end{figure}

\begin{figure}
\plotone{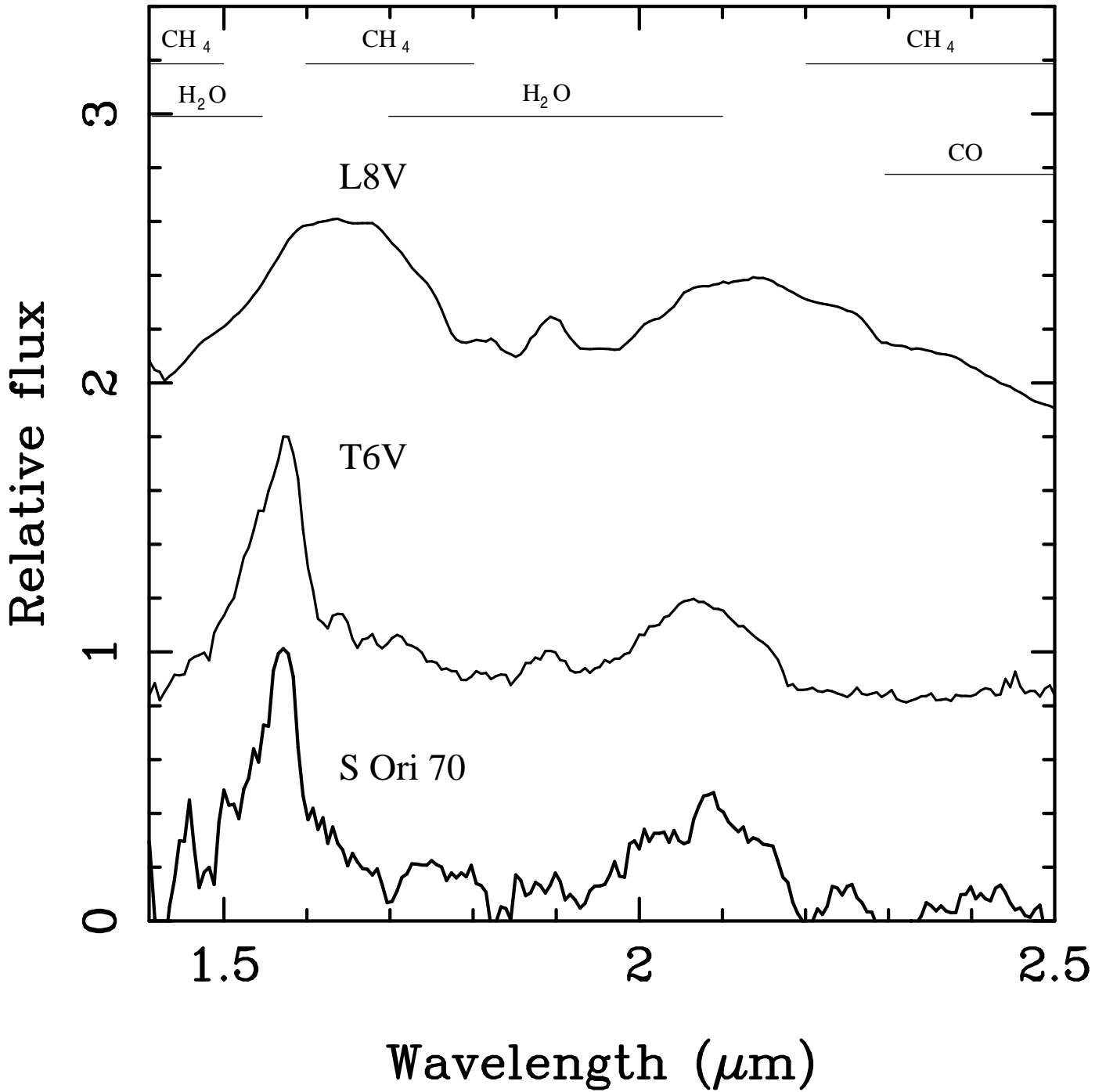}
\caption{NIRC spectrum of S\,Ori\,70 (smoothed with a boxcar of 5
pixel). The two spectral standard field brown dwarfs
SDSS\,J0857585+570851, with spectral type L8V, and
2MASSI\,J0243137--245329, with spectral type T6V, obtained with NIRC
on the same night as S\,Ori\,70, are shown for comparison. All spectra
are normalized to their peak flux in the $H$ band, and have been
shifted by 1 unit for clarity. Most relevant features are
indicated. \label{spectrum}}
\end{figure}

\begin{figure}
\plotone{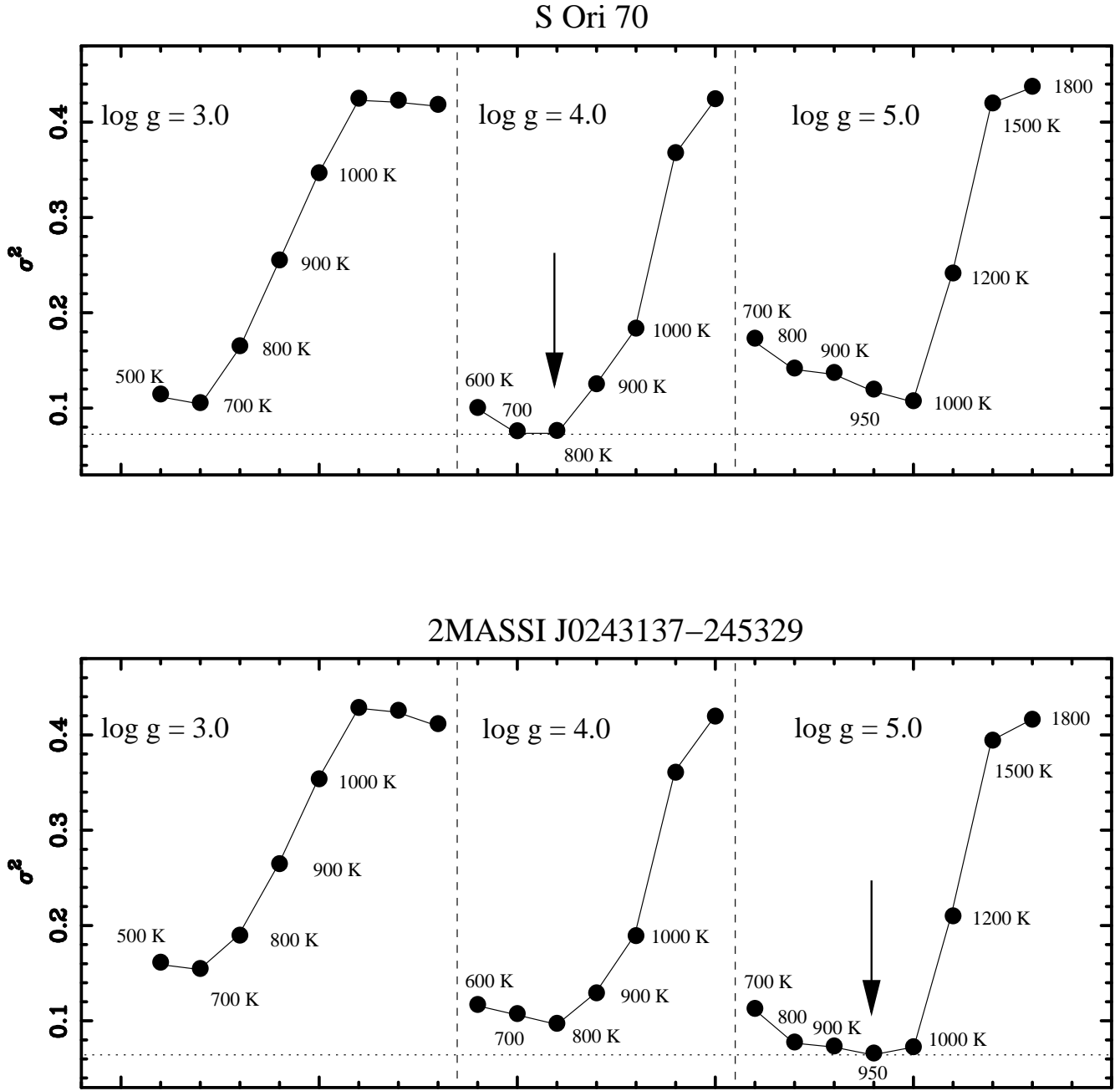}
\caption{Residuals of fits of the observed near-infrared $HK$ spectrum
of S\,Ori\,70 (upper panel) and of the T6V-type field brown dwarf
2MASSI\,J0243137--245329 (lower panel) to synthetic spectra covering a
range of $T_{\rm eff}$ and log\,$g$. The best-fitting models are
marked by arrows for both objects. The synthetic spectra assume solar
composition (Allard et al$.$ \cite{allard01}). Effective temperatures
(K) are indicated. Logarithmic gravities are in
cm\,s$^{-2}$.\label{logg}}
\end{figure}

\begin{figure}
\plottwo{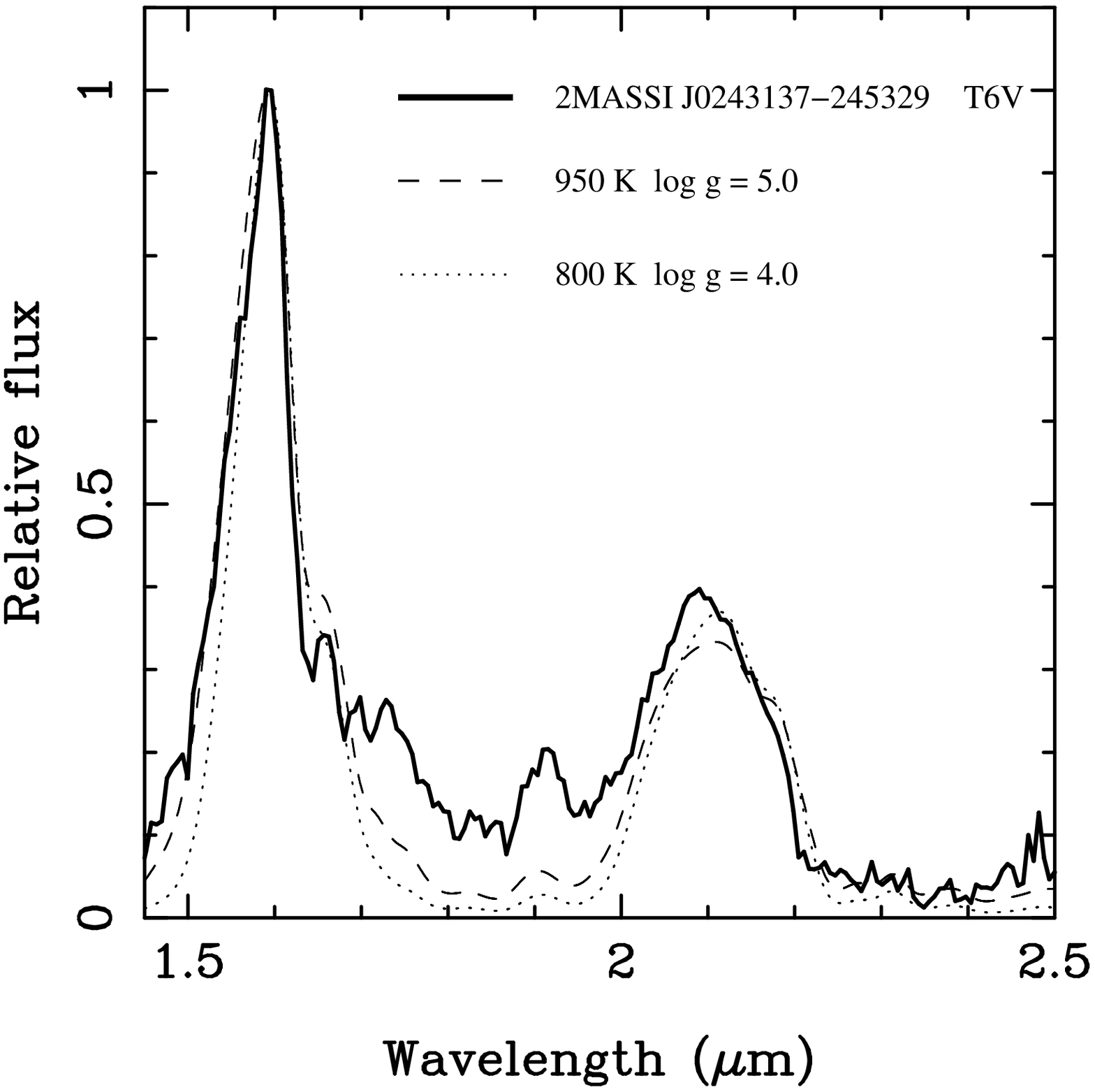}{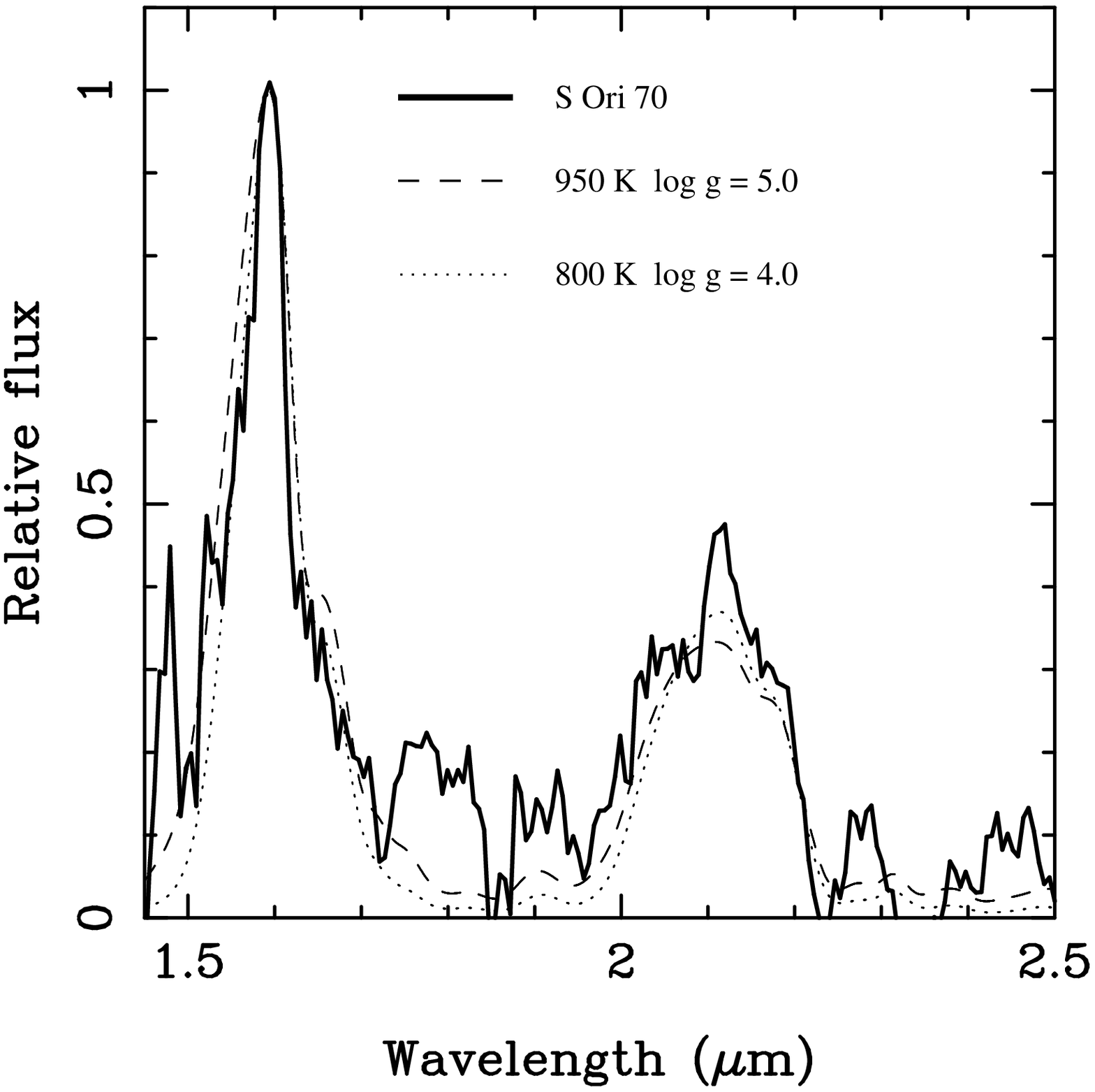}
\caption{Comparison of observed spectra to theoretical {\sc ``cond''}
spectra (Allard et al$.$ \cite{allard01}), which have been degraded to
match the resolution of the observations. Logarithmic gravities are in
cm\,s$^{-2}$. Spectral features and normalization are as in
Figure~\ref{spectrum}.\label{models}}
\end{figure}

In the following analysis, we derive the properties of S\,Ori\,70
assuming that it is a member of the $\sigma$\,Orionis cluster. Its
mass is obtained through comparison with state-of-the-art evolutionary
models.  This exercise is complicated by the marked dependence of
T-class colors on gravity (see Table~\ref{color}). Overplotted onto
the data of Figure~\ref{fotom} are two isochrones for 1 and 10\,Myr
from Baraffe et al$.$ \cite{baraffe02}. Theoretical effective
temperatures and luminosities have been transformed into observables
using temperature calibrations and bolometric corrections (Leggett et
al$.$ \cite{leggett02}; Basri et al$.$ \cite{basri00al}; Reid et al$.$
\cite{reid01a}), which have been calculated for field, high-gravity
dwarfs. As shown in Figure~\ref{fotom}, this is a reasonable approach
for warmer spectral types since colors and bolometric corrections are
similar for different gravities. Allard et al$.$ \cite{allard01},
\cite{allard96} noted that for low surface temperatures (below
1500\,K, i.e., T-types), optical wavelengths and the $K$-band are
strongly affected by the gravity change.  This could explain the
apparent lack of consistency between the tracks and the location of
S\,Ori\,70 in Figure~\ref{fotom}. On the other hand, S\,Ori\,70 may be
a binary, thus appearing overluminous in our Figure. Such an
interpretation cannot be excluded, since Mart\'\i n et al$.$
\cite{martin00} argued that the photometric binary fraction of
Pleiades brown dwarfs is about 25\%, which is similar to the
percentage of binaries found among field brown dwarfs (Reid et al$.$
\cite{reid01b}).

\begin{figure}
\plottwo{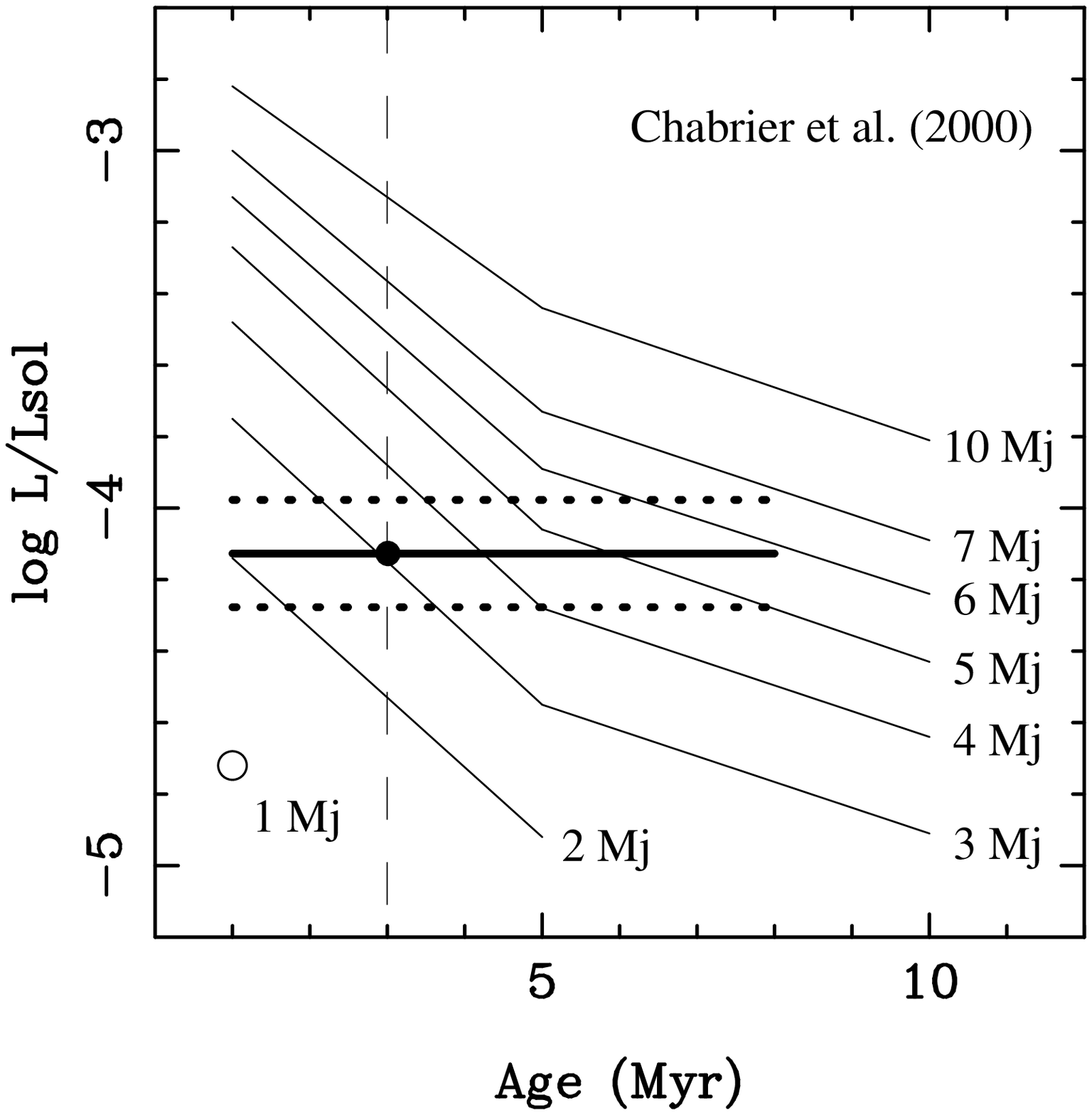}{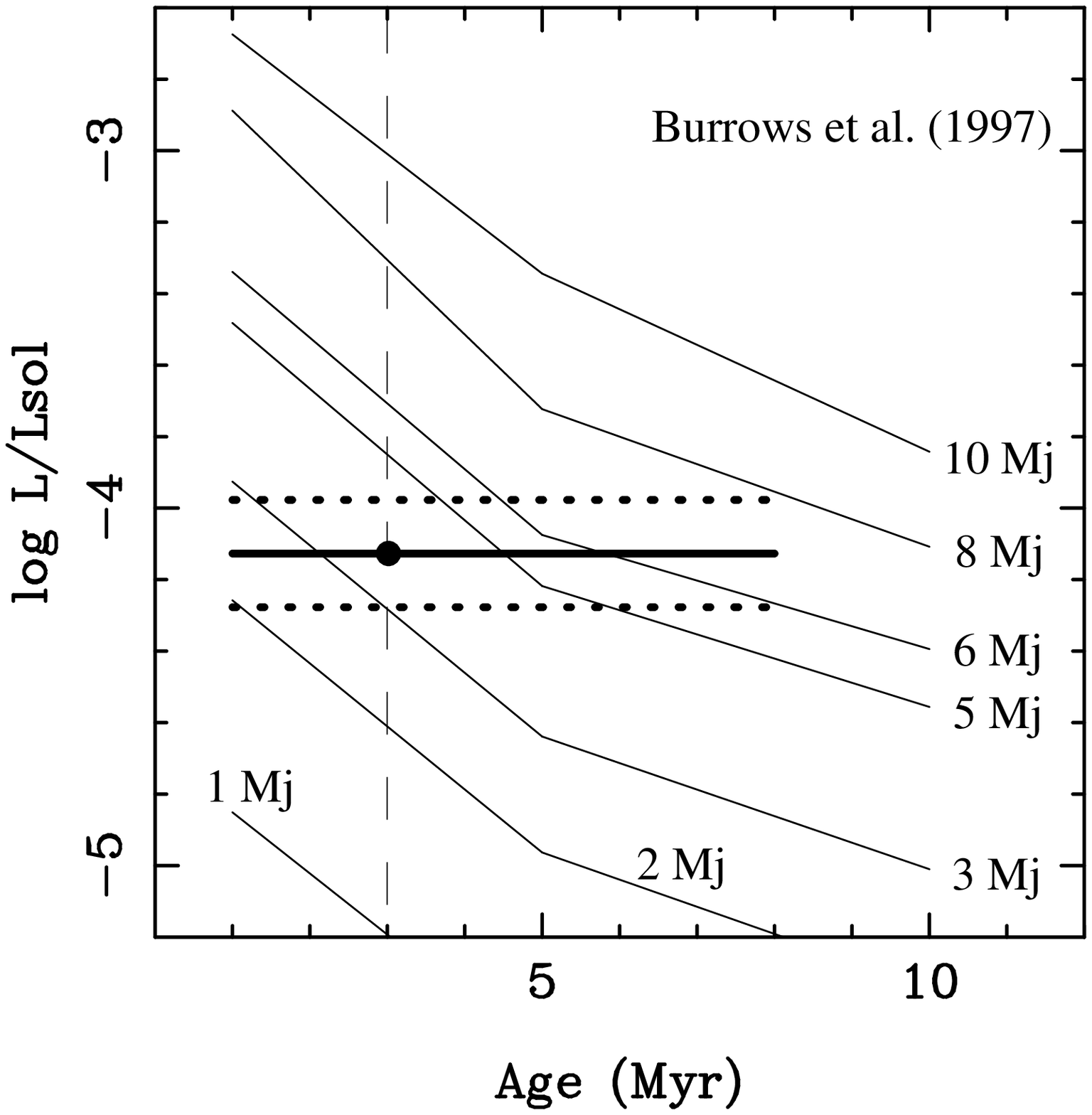}
\caption{Luminosity evolution of very low-mass objects according to
theoretical models by Chabrier et al$.$ \cite{chabrier00} and Burrows
et al$.$ \cite{burrows97}. The location of S\,Ori\,70, considering all
possible cluster ages, is given by the thick line. The most likely
cluster age is estimated at 3\,Myr (vertical dashed line). The
short-dashed lines indicate the error in luminosity. Masses are units
of Jupiter mass.\label{edad}}
\end{figure}

From Table~\ref{color} we infer that the bolometric correction in the
$J$-band is not so much affected by gravity.  This allows us to derive
the luminosity of S\,Ori\,70 to a certain degree of confidence. We
find it to be log\,$L/L_{\odot}$\,=\,--4.13\,$\pm$\,0.15.  For this,
we have used the averaged $J$ bolometric correction of T6V and T8V
field brown dwarfs from Leggett et al$.$
\cite{leggett02}. Figure~\ref{edad} displays the luminosity evolution
of objects of different masses according to the models by Chabrier et
al$.$ \cite{chabrier00} and Burrows et al$.$
\cite{burrows97}. S\,Ori\,70 is indicated with a thick line that takes
into account the age interval adopted for the $\sigma$\,Orionis
cluster (1--8\,Myr). The mass of our ``methane'' object can be derived
from the Figure at 3\,$^{+5}_{-1}$\Mj, where we have adopted 3\,Myr as
the most likely cluster age (Wolk \cite{wolk96}; B\'ejar, Zapatero
Osorio, \& Rebolo \cite{bejar99}; Zapatero Osorio et al$.$
\cite{osorio02}; Oliveira et al$.$ \cite{oliveira02}). The upper and
lower error bars take into account differences between models, the age
range of the cluster, and the uncertainty in luminosity. Nevertheless,
we shall caution that the reliability of this mass estimate relies on
the evolutionary tracks, and thus is subject to possible systematic
effects arising from the description of atmospheres.  Our derived mass
is compatible within uncertainties with the estimated $T_{\rm eff}$
and log\,$g$. In relative terms, if S\,Ori\,70 is a member of the
$\sigma$\,Orionis cluster, it is the least massive object so far
identified in isolation beyond our solar system.

Finding S\,Ori\,70 in the small area covered by our near-infrared
survey might have some impact on our knowledge of the
$\sigma$\,Orionis substellar IMF, despite the dubious reliability
based on a single object. Far from suggesting a decrease in the number
of objects with planetary masses below 5--8\Mj~(Lucas \& Roche
\cite{lucas00}; Muench et al$.$ \cite{muench02}), the discovery of
S\,Ori\,70 is consistent with an extension of the IMF toward very low
masses that may still continue with a slope of $\alpha$\,$\sim$\,--1
(d$N$/d$M$\,$\sim$\,$M^{\alpha}$) in the planetary mass domain. This
represents the natural extension of the brown dwarf IMF given in
B\'ejar et al$.$ \cite{bejar01} for the $\sigma$\,Orionis cluster.
However, we remark that our survey only imaged a very small fraction
(0.2\%) of the total cluster area, and that {\it one} object has very
low statistical meaning. Hence, larger area searches need to be
conducted for a better knowledge of the $\sigma$\,Orionis planetary
IMF.

The detection of S\,Ori\,70 indicates that objects only slightly
heavier than Jupiter may exist free-floating in
$\sigma$\,Orionis. Their formation process is not yet
established. Theory predicts that opacity-limited fragmentation of
cool gravitationally collapsing gas clouds is capable of producing
7--10\Mj~Population~I objects in isolation (Silk \cite{silk77}; Low \&
Lynden-Bell \cite{low76}). Moreover, this minimum Jeans mass seems to
be insensitive to changes in the opacity of protostellar clouds
(amount of dust, size of grains, cosmic ray flux; Boss
\cite{boss88}). These models, however, do not include rotation,
magnetic fields, and further external accretion onto the cloud
fragment, which might alter the final mass of the nascent
object. S\,Ori\,70 is probably less massive than the minimum Jeans
mass of 7--10\Mj, and thus prompts us to refine the
collapse-and-fragmentation models and/or to rethink possible formation
mechanisms for such low-mass objects. Recently, several formation
scenarios have been suggested that include tidal interactions and
ejection of low-mass objects from multiple systems before brown dwarfs
and planetary-mass objects can accrete enough gas to become stars
(e.g., Reipurth \& Clarke \cite{reipurth01}; Boss \cite{boss01}; Bate,
Bonnell, \& Bromm \cite{bate02}). S\,Ori\,70 is located 8\arcmin.65 to
the west of the multiple stellar system $\sigma$\,Orionis, which is
formed by six hot massive stars (Morrell \& Levaton \cite{morrell91};
Bolton \cite{bolton74}). S\,Ori\,70 might have been ejected from it
with a space velocity of
$v$\,sin\,$\gamma$\,$\sim$\,0.29\,km\,s$^{-1}$ for a cluster age of
3\,Myr, where $\gamma$ stands for the ejection direction with respect
to the line of sight. On the other hand, Padoan \& Nordlund
\cite{padoan02}, and references therein, suggest that brown dwarfs are
formed in the same way as more massive hydrogen burning stars, that is
by the process of supersonic turbulent fragmentation. These authors
claim that their model for the IMF predicts a number of substellar
objects consistent with observational data. If any or all of these
scenarios are viable solutions to the formation problem can only be
decided when more such low-mass objects are known, and their
properties have been studied in detail.

\acknowledgments We are grateful to F$.$ Allard and I$.$ Baraffe for
providing us computer-ready files of their models. We are also
grateful to the anonymous referee for valuable comments. This paper is
based on observations obtained at the WHT, operated by the Isaac
Newton Group of Telescopes, funded by PPARC at the Spanish
Observatorio del Roque de los Muchachos of the Instituto de Astrof\'\i
sica de Canarias; at the W.\,M$.$ Keck Observatory, which is operated
as a scientific partnership among the California Institute of
Technology, the University of California and the National Aeronautics
and Space Administration (the Observatory was made possible by the
generous financial support of the W.\,M. Keck Foundation); and at the
Carlos S\'anchez Telescope, at the Spanish Teide Observatory of the
Instituto de Astrof\'\i sica de Canarias. This publication makes use
of data products from the Two Micron All Sky Survey, which is a joint
project of the University of Massachusetts and the Infrared Processing
and Analysis Center/California Institute of Technology, funded by the
National Aeronautics and Space Administration and the National Science
Foundation. Partial financial support comes from the Spanish
``Programa Ram\'on y Cajal''.

\clearpage

\clearpage

\clearpage

\clearpage

\clearpage

\clearpage

\clearpage
\begin{deluxetable}{cccccccccccc}
\tabletypesize{\scriptsize}
\tablecaption{Data for S\,Ori\,70. \label{table}}
\tablewidth{0pt}
\tablehead{
\colhead{        } & 
\colhead{RA} & 
\colhead{DEC} &
\colhead{   } &
\colhead{     } & 
\colhead{     } & 
\colhead{     } & 
\colhead{     } &
\colhead{$T_{\rm eff}$} & 
\colhead{log\,$g$} &
\colhead{   }                  \\
\colhead{IAU name} & 
\colhead{(J2000)} & 
\colhead{(J2000)} &
\colhead{$J$} &
\colhead{$J-H$} & 
\colhead{$J-K_s$} & 
\colhead{$I-J$} &
\colhead{$I-z$} &
\colhead{(K)} & 
\colhead{(cm\,s$^{-2}$)} &
\colhead{SpT}
}
\startdata
S\,Ori\,J053810.1--023626 & 05 38 10.1 & --02 36 26 & 
20.28$\pm$0.10 & --0.14$\pm$0.15 & 0.50$\pm$0.20 & 
4.75$\pm$0.40 & 2.2$\pm$0.4 & 700--1000 & 4.0$\pm$1.0 & T class\\
\enddata
\tablecomments{Units of right ascension are hours, minutes, and
seconds, and units of declination are degrees, arcminutes, and
arcseconds. Coordinates are accurate to $\pm$2\arcsec. See text 
for the calibration of the $z$-band.}
\end{deluxetable}

\begin{deluxetable}{cccccc}
\tabletypesize{\scriptsize}
\tablecaption{Color differences \label{color}}
\tablewidth{0pt}
\tablehead{
\colhead{} & 
\colhead{     } & 
\colhead{S\,Ori\,70} & 
\colhead{S\,Ori\,70} & 
\colhead{S\,Ori\,70} \\ 
\colhead{} & 
\colhead{Model\tablenotemark{a}} & 
\colhead{-- T6V} & 
\colhead{-- T7V} &
\colhead{-- T8V} 
}
\startdata
$I-J$ & $-$0.47 & $-$0.9  & \nodata & \nodata \\
$J-H$ & $-$0.10 & $-$0.21 & $+$0.04 & $-$0.23 \\
$H-K$ & $+$0.49 & $+$0.83 & $+$0.31 & $+$0.59 \\
$J-K$ & $+$0.38 & $+$0.53 & $+$0.35 & $+$0.36 \\
\enddata
\tablenotetext{a}{{\sc ``cond''} 900\,K, 
  [log\,$g$\,=\,3.0]\,--\,[log\,$g$\,=\,5.0]}
\end{deluxetable}

\end{document}